\documentclass[preprint,showpacs,preprintnumbers,amsmath,amssymb]{revtex4}

\usepackage{graphicx}
\usepackage{graphics}
\usepackage{subfigure}
\usepackage{dcolumn}
\usepackage{bm}
\usepackage{txfonts}

\begin{document}

\preprint{}

\title{Effects of empathy on the evolution of fairness in group-structured populations}

\author{Yanling Zhang$^{1\ddagger}$}
\author{Jian Liu$^{1\ddagger}$}
\author{Aming Li$^{2}$}
\email{aming.li@zoo.ox.ac.uk}

\affiliation{$^1$Key Laboratory of Knowledge Automation for Industrial Processes of Ministry of Education, School of Automation and Electrical Engineering, University of Science and Technology Beijing, Beijing 100083, China\\
$^2$Department of Zoology, University of Oxford, Oxford OX13PS, United Kingdom\\
$\ddagger$ These authors contributed equally to this work.}

\date{\today}

\begin{abstract}
The ultimatum game has been a prominent paradigm in studying the evolution of fairness. It predicts that responders should accept any nonzero offer and proposers should offer the smallest possible amount according to orthodox game theory. However, the prediction strongly contradicts with experimental findings where responders usually reject low offers below $20\%$ and proposers usually make higher offers than expected. To explain the evolution of such fair behaviors, we here introduce empathy in group-structured populations by allowing a proportion $\alpha$ of the population to play empathetic strategies.
Interestingly, we find that for high mutation probabilities, the mean offer decreases with $\alpha$ and the mean demand increases, implying empathy inhibits the evolution of fairness.
For low mutation probabilities, the mean offer and demand approach to the fair ones with increasing $\alpha$, implying empathy promotes the evolution of fairness.
Furthermore, under both weak and strong intensities of natural selection, we analytically calculate the mean offer and demand for different levels of $\alpha$.
Counterintuitively, we demonstrate that although a higher mutation probability leads to a higher level of fairness under weak selection, an intermediate mutation probability corresponds to the lowest level of fairness under strong selection.
Our study provides systematic insights into the evolutionary origin of fairness in group-structured populations with empathetic strategies.
\end{abstract}


\maketitle

\section{Introduction}
In the canonical ultimatum game, one player proposes a division of a sum of money between himself/herself and a second player, who might accept or reject the proposal.
If the proposal is accepted, the sum is shared accordingly;
if not, both players remain empty handed.
On the assumption of orthodox game theory, the responder should accept any nonzero offer and the proposer should offer the smallest possible amount.
However, human behavioral experiments show a different reality: responders reject low offers below $20\%$ about half of the time and proposers usually make higher offers than expected to avoid rejection~\cite{guth1982experimental,camerer2011behavioral,guth2014more}.
The problem on the evolution of fairness has been mainly studied by a great diversity of models under the framework of game theory and evolutionary game theory~\cite{debove2016models}.

Evolutionary game dynamics focuses on the evolution of fairness among selfish and interactive individuals~\cite{szolnoki2018reciprocity,wang2018replicator,perc2017statistical,yang2018promoting,perc2016phase,szolnoki2017second,li2015evolutionary,perc2015double,fu2013global,liu2017randomness,liu2018fixeddelay}, where strategies with higher fitness are more likely to spread among individuals. Under this evolutionary framework, the conditions under which the fair strategy is favored by natural selection have been attracting considerable interest. In unstructured populations without invoking additional mechanisms, natural selection itself can lead to the rational self-interest strategy, where agents offer the smallest amount and accept any nonzero offer~\cite{nowak2000fairness, page2001generalized}.
Considering this, researchers have investigated a large number of factors to explain how the fair strategy evolves in the two-person ultimatum game~\cite{nowak2000fairness, page2001generalized,sanchez2005altruism,szolnoki2012defense,sinatra2009ultimatum,page2002empathy,wu2013adaptive, yang2015effects,rand2013evolution,wang2014random,forber2014evolution,zhang2018strategy,zhang2018effect,szolnoki2012accuracy,page2000spatial, killingback2001spatial,kuperman2008effect,gao2011coevolutionary,iranzo2011spatial,wang2015evolutionary,liu2018fixedtime} as well as the multi-person scenario~\cite{takesue2017evolution, nishimura2017evolution}. For the two-person ultimatum game, these factors are reputation~\cite{nowak2000fairness}, empathy~\cite{page2001generalized,szolnoki2012defense,page2002empathy,sinatra2009ultimatum,sanchez2005altruism}, alternating role~\cite{wu2013adaptive, yang2015effects},
randomness~\cite{rand2013evolution,wang2014random}, spite~\cite{forber2014evolution,zhang2018strategy},
altering the stake size~\cite{zhang2018effect},
spatial population structure~\cite{szolnoki2012defense,szolnoki2012accuracy,page2000spatial, killingback2001spatial,kuperman2008effect,gao2011coevolutionary,iranzo2011spatial,wang2015evolutionary,liu2018fixedtime}, and so on.
Particularly, some models comprise different factors, for example, the effects of empathy~\cite{szolnoki2012defense}, adaptive role switching~\cite{wu2013adaptive, yang2015effects}, random allocation~\cite{wang2014random}, and migration~\cite{wang2015evolutionary}
have been studied in spatially structured populations.

Empathy, meaning that individuals only make offers that they would themselves be ready to accept,
has attracted much attention in modeling human behavior. When all players are assumed to play empathetic strategies, fair offers can be evolved in unstructured populations~\cite{sanchez2005altruism}.
Under the same assumption, empathetic players also result in the evolution of fairness in regular graphs~\cite{szolnoki2012defense} and other complex networks~\cite{sinatra2009ultimatum}.
Page and Nowak have found that in unstructured populations, allowing a small proportion $\alpha$ of the population to play empathetic strategies is enough to favor the evolution of fairness~\cite{page2002empathy}.
However, if natural selection acts upon $\alpha$, they also have shown that empathy is not selected by natural selection.
Considering this, Iranzo et al. have reported that empathy itself can be selected whenever players are restricted to interact with a fraction of the total population~\cite{iranzo2012empathy}.

Here, we study the effects of empathy on the evolution of fairness in group-structured populations.
Indeed, the group-structured population is ubiquitous in nature~\cite{li2016evolutionary,santos2015evolutionary,iranzo2012empathy}, for example, the group can be understood as an island in population genetics or a particular organization in human society.
We introduce empathy by initially allowing a proportion $\alpha$ of the population to play empathetic strategies. It is noteworthy that the single-group case of our model degenerates to the simple scenario~\cite{page2002empathy}, where authors have only focused on small $\alpha$ and very low mutation probabilities.
In contrast, we make a comprehensive analysis here by varying $\alpha$ and the mutation probability from $0$ to $1$. Furthermore, we also investigate how the intensity of selection influences the evolution of fairness.
Considering the advantages of temporal networks over their static counterparts~\cite{li2017fundamental,rodrigues2016kuramoto,saavedra2009simple,masuda2018configuration}, we introduce migration in the model and allow the interactive network to change with time.
Under both weak and strong intensities of selection, we analytically provide the mean offer and demand for different levels of $\alpha$.

\section{Model}
Consider a population with $N$ players, in which each player belongs to one of $M$ groups.
 Each player (say $i$) interacts with all other players in the same group, and receives the total payoff $P_i$. The fitness of the player $i$ is defined as $1-\omega+\omega P_i$, where $\omega$ is called the intensity of natural selection~\cite{Nowak2004Emergence}. The extreme case $\omega\rightarrow 0$ corresponds to weak selection. For $\omega=1$, the fitness is identical to the payoff, and the case $\omega\rightarrow 1$ corresponds to strong selection. When two players interact, they play the ultimatum game twice, and they are assigned to the roles of proposer and responder in turn. The proposer suggests how to split a fixed sum (say 1) and the responder decides whether or not to accept the proposal. When the responder accepts it, the offer is split accordingly; otherwise, both players get nothing. The strategy is usually denoted by a vector $(p,q)$ where $p\in[0,1]$ is the amount offered when the player acts as a proposer, and $q\in[0,1]$ is the minimum amount demanded when the player acts as a responder. An offer $p_1$ is accepted by a responder with the minimum demand $q_2$ if and only if $p_1\geq q_2$. Therefore, the payoff for a player using the strategy $(p_1,q_1)$ against another player using the strategy $(p_2,q_2)$ is given by
\begin{eqnarray}\label{1}
a((p_1,q_1),(p_2,q_2))=\left\{
\begin{array}{ll}
0 & p_1<q_2,p_2<q_1\\
p_2 &p_1< q_2,p_2\ge q_1\\
1-p_1 & p_1\ge q_2,p_2 <q_1\\
1-p_1+p_2 & p_1\geq q_2,p_2\geq q_1\\
\end{array}\right..
\label{payoff}
\end{eqnarray}

\begin{figure}[t]
\centerline{\includegraphics[width=0.8\textwidth]{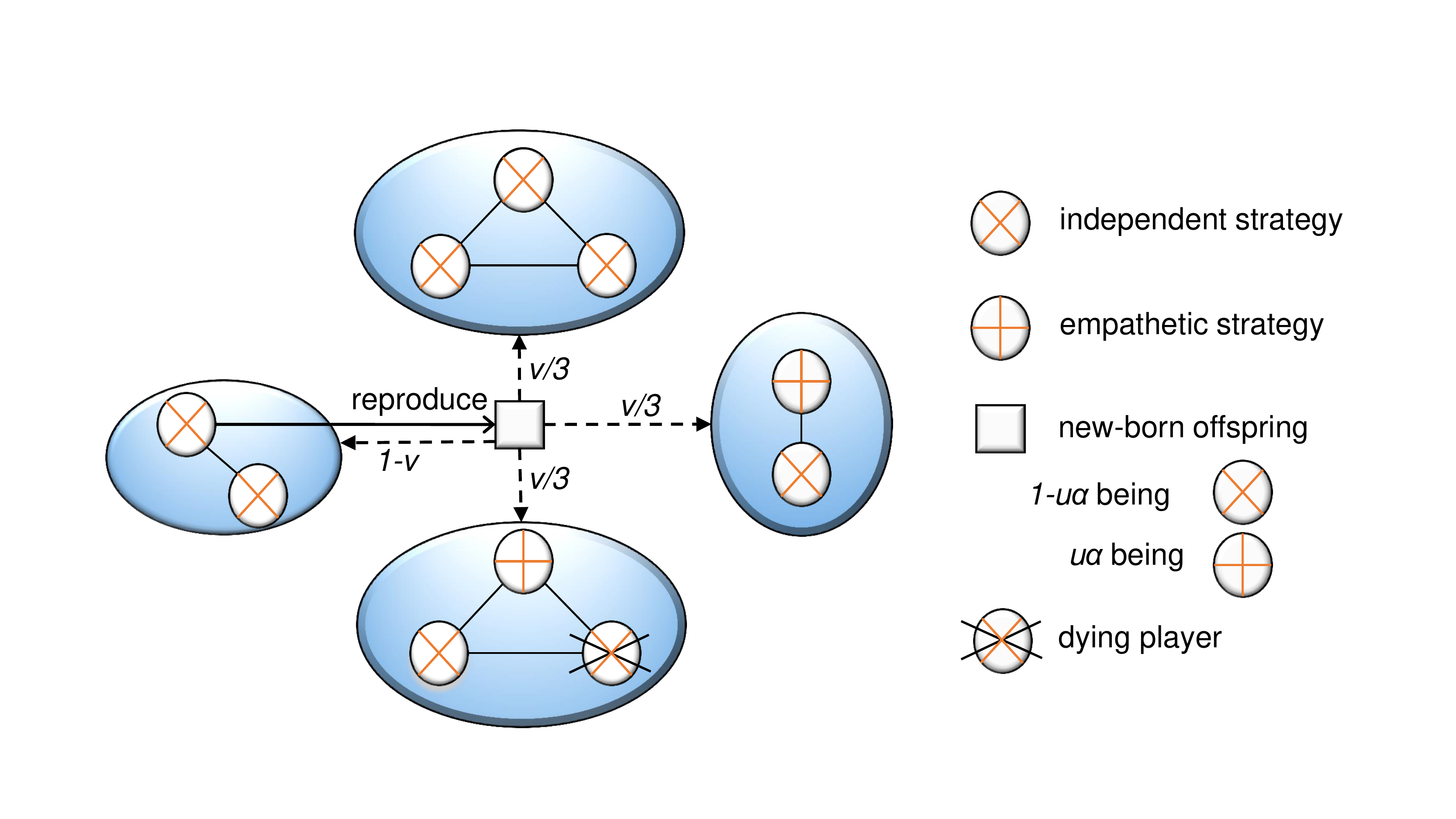}}
\caption{(Color online) Model schematic. Ten players (represented by nodes, $N=10$), eight of whom (nodes with $\times$) play independent strategies and the remainder (nodes with $+$) play empathetic strategies ($p=q$), are distributed over four groups (shaded by blue, $M=4$). In a given generation, a player with an independent strategy is chosen to reproduce one offspring (empty square) and a player (node under cross) is chosen to die. The newborn offspring randomly adopts an independent strategy with probability $1-u\alpha$ and an empathetic strategy with $u\alpha$. The offspring stays in his parent's group with $1-v$ and he randomly migrates to one other group with $v$.}
\label{fig1}
\end{figure}
If a strategy $(p,q)$ satisfies $p=q$, i.e., the offer that a player makes is equal to the minimum offer that he prepares to accept, the strategy is called as an empathetic strategy.
If $p$ and $q$ of a strategy $(p,q)$ are independent of each other, the strategy is an independent strategy. Initially, each player adopts a random empathetic strategy with probability $\alpha\in[0,1]$, and adopts a random independent strategy with probability $1-\alpha$.
The update follows the frequency-dependent Moran process~\cite{Nowak2004Emergence}. As shown in Fig.~\ref{fig1}, a player is chosen proportional to his fitness to reproduce one offspring and a player is randomly chosen to die in each generation. The newborn offspring adopts his parent's strategy with probability $1-u$; otherwise, he mutates to a random empathetic strategy with probability $\alpha$ and a random independent strategy with probability $1-\alpha$. Moreover, the newborn offspring is allowed to migrate to one new group with probability $v$; otherwise, he stays in his parent's group. We assume that the $M$ groups are located in a circle. The new group can be one of two groups closest to his parent's group for local migration, and any one group except his parent's group for global migration. It is noteworthy that all players adopt independent strategies for $\alpha=0$ and empathetic strategies for $\alpha=1$ during the whole evolutionary process.
For $0<\alpha<1$, the fraction of empathetic strategies fluctuates around $\alpha$ during the evolutionary process. For simplicity, $\alpha$ is called the fraction of empathetic strategies hereafter.

\section{Results}
We first perform agent-based simulations by varying the fraction of empathetic strategies $\alpha$ and the intensity of natural selection $\omega$. For each set of simulation parameters, we determine the mean offer and demand, which are the time-averaged values of $p$ and $q$ over the whole population, respectively.

\begin{figure}[t]
\centerline{\includegraphics[width=0.8\textwidth]{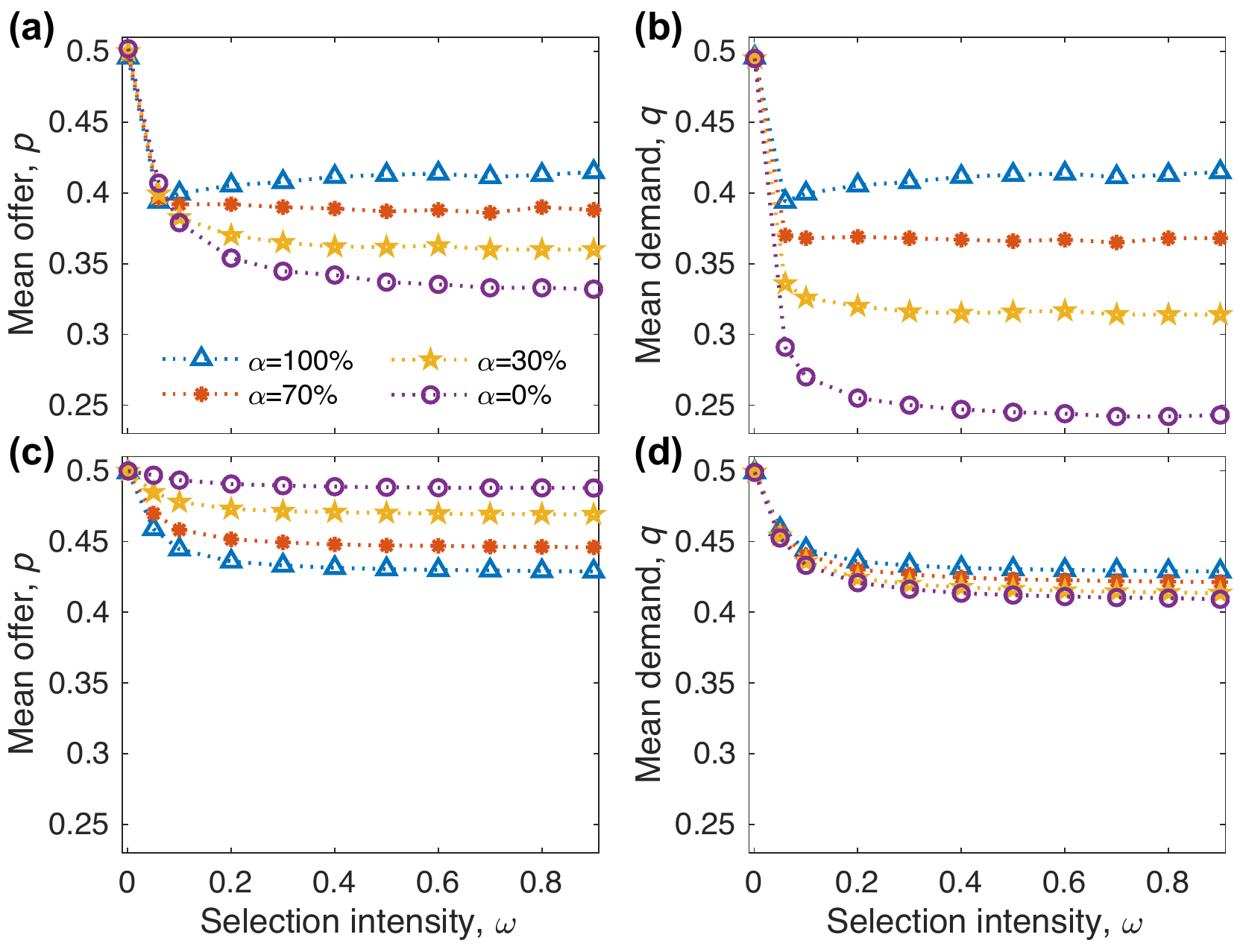}}
\caption{(Color online) The changing trends of the mean offer $p$ and demand $q$ with the intensity of selection $\omega$. $p$ and $q$ significantly decrease when $\omega$ increases from $0$. They remain around non-zero constants when $\omega$ is sufficiently large. ((a), (b)) For low mutation probabilities ($u=0.1$), $p$ and $q$ approach to $0.5$ with increasing fraction of empathetic strategies $\alpha$. ((c), (d)) For high mutation probabilities ($u=0.4$), $p$ decreases with $\alpha$ and $q$ increases with $\alpha$. Parameters: $N =50$, $M = 9$, $r =1$, and $v = 0.1$.}
\label{fig2}
\end{figure}
Figure~\ref{fig2} shows how $\omega$ influences the mean offer and demand. When $\omega$ is very small ($\omega\approx0$), neutral drift dominates the evolutionary dynamics, indicating that the mean offer and demand are close to $0.5$. As $\omega$ increases but is still small, both the mean offer and demand decrease significantly with $\omega$.
When $\alpha$ is small, the decreasing trends mainly depend on the evolution of independent strategies. According to Eq.~(\ref{payoff}), a lower offer has a payoff advantage over a higher offer for independent strategies, i.e., $a((p,*),(p+\epsilon,*))\geq a((p+\epsilon,*),(p,*))$ with $\epsilon>0$. Moreover, the increase of $\omega$ significantly enlarges the payoff advantage of lower offers over higher offers. Therefore, we have the significant decrease of the mean offer with $\omega$.
The decrease of the mean offer drives the mean demand to decrease accordingly in order to avoid rejection. When $\alpha$ is large, the decreasing trends mainly depend on the evolution of empathetic strategies. According to Eq.~(\ref{payoff}), for empathetic strategies, the strategy with $p<0.5$ has a payoff advantage over the strategy with $p>0.5$, i.e., the payoff of $p=0.5-\epsilon$ ($\epsilon>0$) is more than that of $p=0.5+\epsilon$. The initial increase of $\omega$ enlarges the payoff advantage of $p<0.5$ over $p>0.5$, and it leads to the significant decrease of the mean offer and demand. When $\omega$ becomes sufficiently large and further increases, the mean offer or demand remains around non-zero constants. The reason for the existence of the non-zero constant is as follows. The numerator and the denominator of the probability that a player reproduces an offspring linearly depend on $\omega$. When $\omega$ is sufficiently large, it can be simultaneously eliminated from the numerator and the denominator. Accordingly, the payoff advantage of lower offers over higher offers or the payoff advantage of $p<0.5$ over $p>0.5$ cannot be enlarged by increasing $\omega$. Consequently, the mean offer or demand can not change with $\omega$ and remains around non-zero constants.

\begin{figure}[h]
	\centerline{\includegraphics[width=0.8\textwidth]{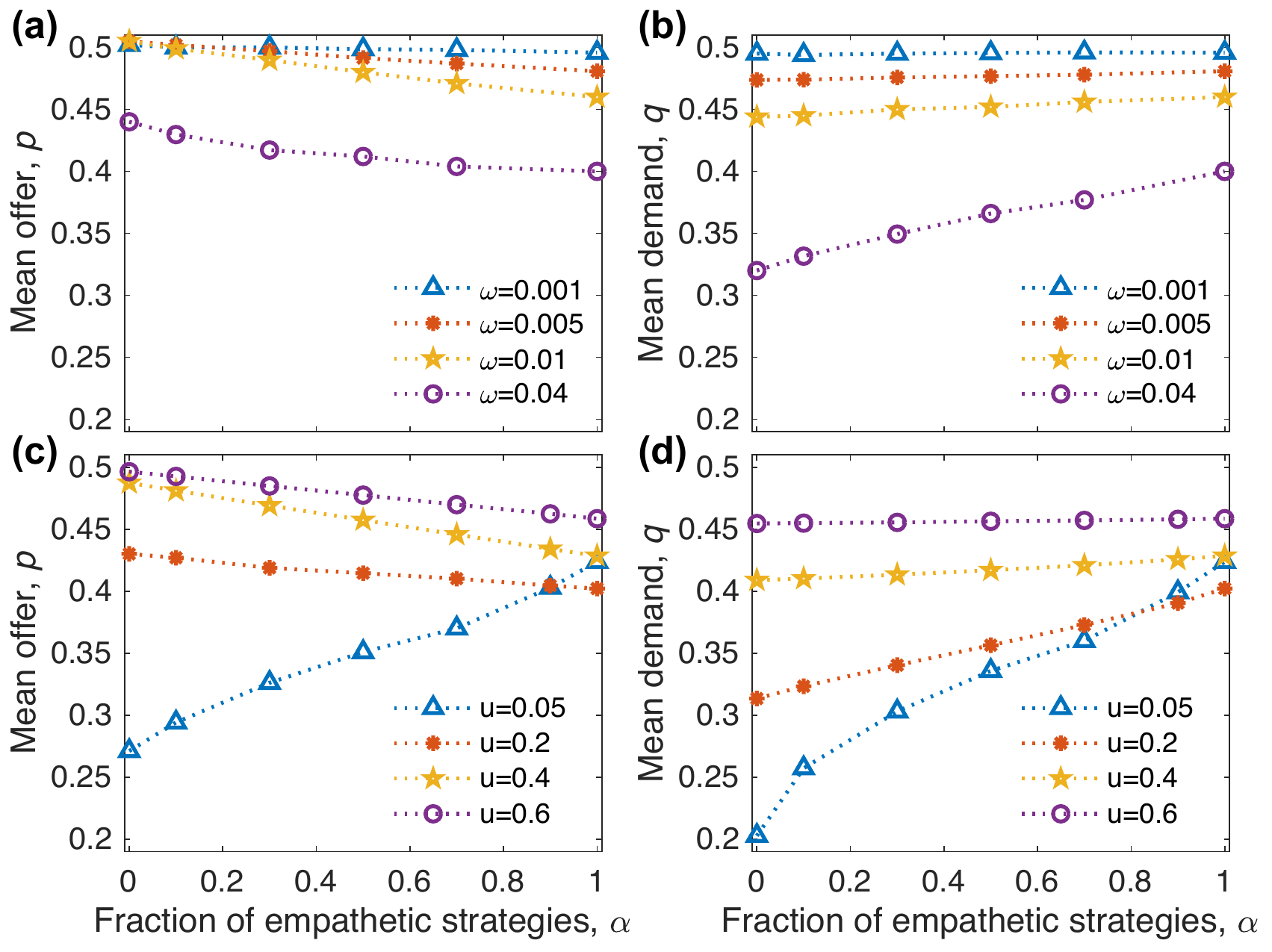}}
	\caption{(Color online) The changing trends of the mean offer $p$ and demand $q$ with the fraction of empathetic strategies $\alpha$.
$p$ and $q$ change very little with $\alpha$ for sufficiently small intensities of selection $\omega$ in ((a), (b)) and sufficiently high mutation probabilities $u$ in ((c), (d)).
(c) $p$ increases with $\alpha$ for low $u$, but decreases with $\alpha$ for high $u$.
(d) $q$ increases with $\alpha$ for all $u$.
Parameters: $N =50$, $M = 9$, $r =1$, $v = 0.1$, ((a), (b)) $u=0.1$, and ((c), (d)) $\omega=1$.}
\label{fig3}
\end{figure}
As shown in Figs.~\ref{fig2} and~\ref{fig3}, the effects of $\alpha$ on the mean offer and demand depend on $\omega$ and the mutation probability $u$. When $\omega$ is very small, the mean offer and demand change very little with $\alpha$ (Figs.~\ref{fig3}a,~\ref{fig3}b), because neutral drift dominates the evolutionary dynamics. As sufficiently large $\omega$ further increases, the changing trends of the mean offer and demand with $\alpha$ are affected by $u$.
When $u$ is low, they both approach to the fair ones with increasing $\alpha$ (Figs.~\ref{fig2}a,~\ref{fig2}b,~\ref{fig3}c,~\ref{fig3}d), implying that empathy promotes the evolution of fairness.
When $u$ is high, the mean offer decreases with $\alpha$, but the mean demand increases with $\alpha$ (Figs.~\ref{fig2}c,~\ref{fig2}d,~\ref{fig3}c,~\ref{fig3}d), implying that empathy inhibits the evolution of fairness.
Particularly, they both change very little when $u$ is sufficiently high (Figs.~\ref{fig3}c,~\ref{fig3}d). The results can be intuitively understood as follows. When $u$ is low, the payoff difference plays the major role in the strategy selection compared to the mutation process.
We take two extreme cases $\alpha=0$ and $\alpha=1$ for instance.
We have known that lower offers have the payoff advantage over higher offers for $\alpha=0$. According to Eq.~(\ref{payoff}), the offer closer to $0.5$ has the payoff advantage over the one further away $0.5$ for $\alpha=1$, i.e., a strategy with $p>0.5$ has a higher payoff than that with $p+\epsilon$ ($\epsilon>0$) and a strategy with $p<0.5$ has a higher payoff than that with $p-\epsilon$ ($\epsilon>0$).
Therefore, the mean offer for $\alpha=0$ is less than the one for $\alpha=1$.
Accordingly, the mean demand for $\alpha=0$ is less than the one for $\alpha=1$ in order to avoid rejection. When $u$ is high, the mutation process plays the major role in the strategy selection compared to the payoff difference.
By means of the mutation process, the strategies with high offers or high demands are frequently introduced in the evolutionary process for $\alpha=0$, but only the strategies with high offers and high demands are frequently introduced for $\alpha=1$.
However, only the strategies with high offers and low demands can achieve enough payoffs to survive in the population.
Therefore, the mean offer for $\alpha=0$ is more than the one for $\alpha=1$, but the mean demand for $\alpha=0$ is less than the one for $\alpha=1$.

\begin{figure}[t]
\centerline{\includegraphics[width=0.7\textwidth]{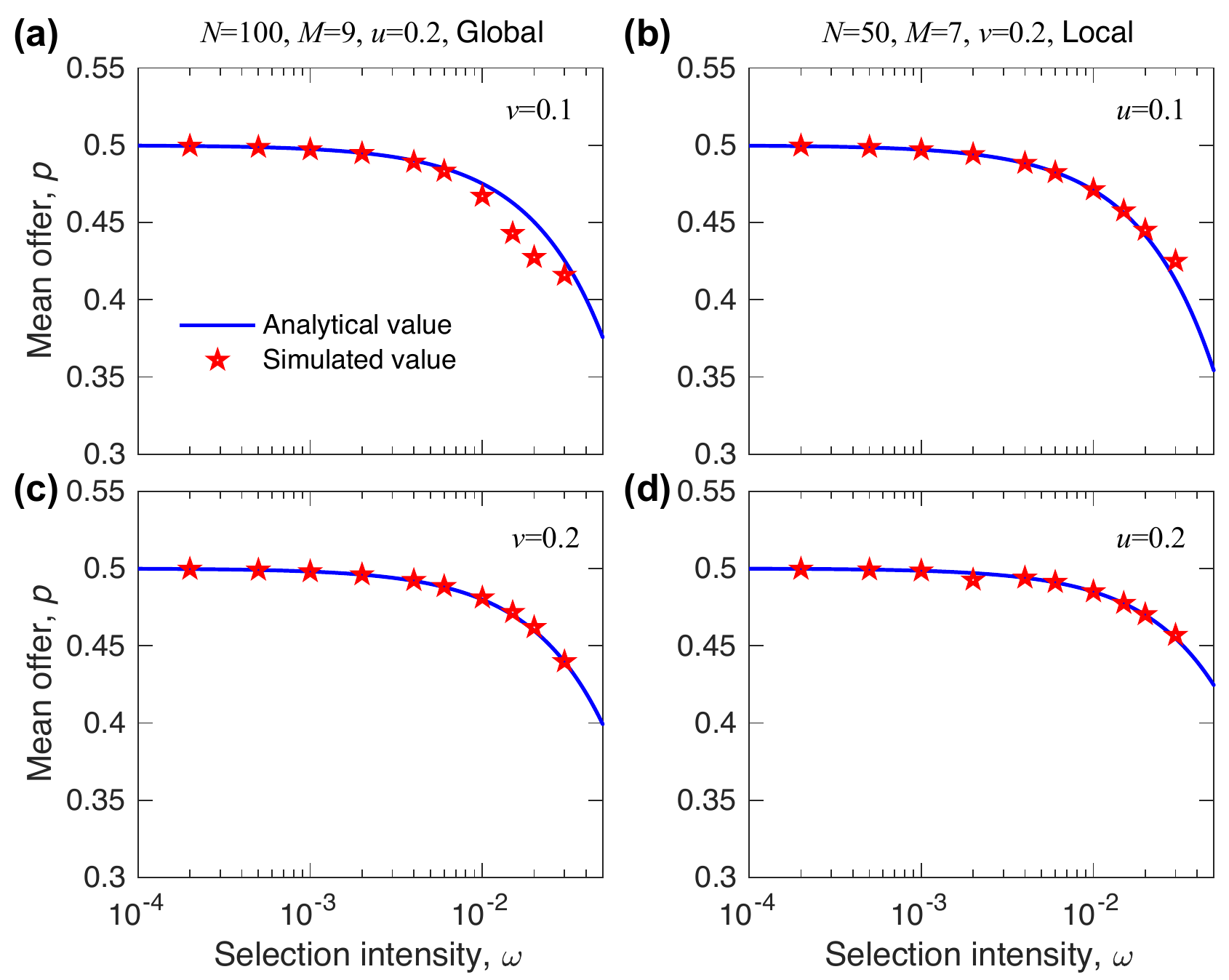}}
	\caption{(Color online) Under weak selection, comparison of theoretical results and numerical results.
When the intensity of selection $\omega$ is small, our analytical calculations of the mean offer $p$ are in good agreement with the simulated values for different population sizes ($N$), group numbers ($M$), mutation probabilities ($u$), migration probabilities ($v$), and migration patterns (both local and global migration).}
\label{fig4}
\end{figure}
After presenting agent-based numerical results, we turn to theoretical calculations.
It has been shown that under weak selection, the mean offer and demand change little with $\alpha$. Accordingly, we focus on the case with $\alpha=1$, where all players use continuous empathetic strategies. Here, a strategy can be described by a single parameter $p\in[0,1]$. The mean offer $p$ is calculated in the Appendix, and its analytical expression is as follows:
\begin{eqnarray}
p=\frac{1}{2}-\omega\frac{(1-u)(N-1)}{{24Mu}}\sum_{x=1}^M(\Psi_1(f(x))-\Psi_2(f(x))),
\label{empathy}		
\end{eqnarray}
where $\Psi_1(f)=\frac{1-v(1-f)}{1+(N-1)v(1-f)}$, $\Psi_2(f)=\frac{(1-u)(1-v(1-f))}{1+(N-1)u+(N-1)(1-u)v(1-f)}$, and $f(x)$ describes the migration pattern.
The expression of $f(x)$ is $f(x)=\frac{1}{M-1}\sum_{j=1}^{M-1}\cos\frac{2\pi j x}{M}$ for global migration and is $f(x)=\cos\frac{2\pi  x}{M}$ for local migration.
Figure~\ref{fig4} shows when the intensity of selection is small, analytical results agree well with numerical results for different population sizes ($N$), mutation probabilities ($u$), migration probabilities ($v$), group numbers ($M$), and migration patterns (both local and global migration).

\begin{figure}[t]
\centerline{\includegraphics[width=0.8\textwidth]{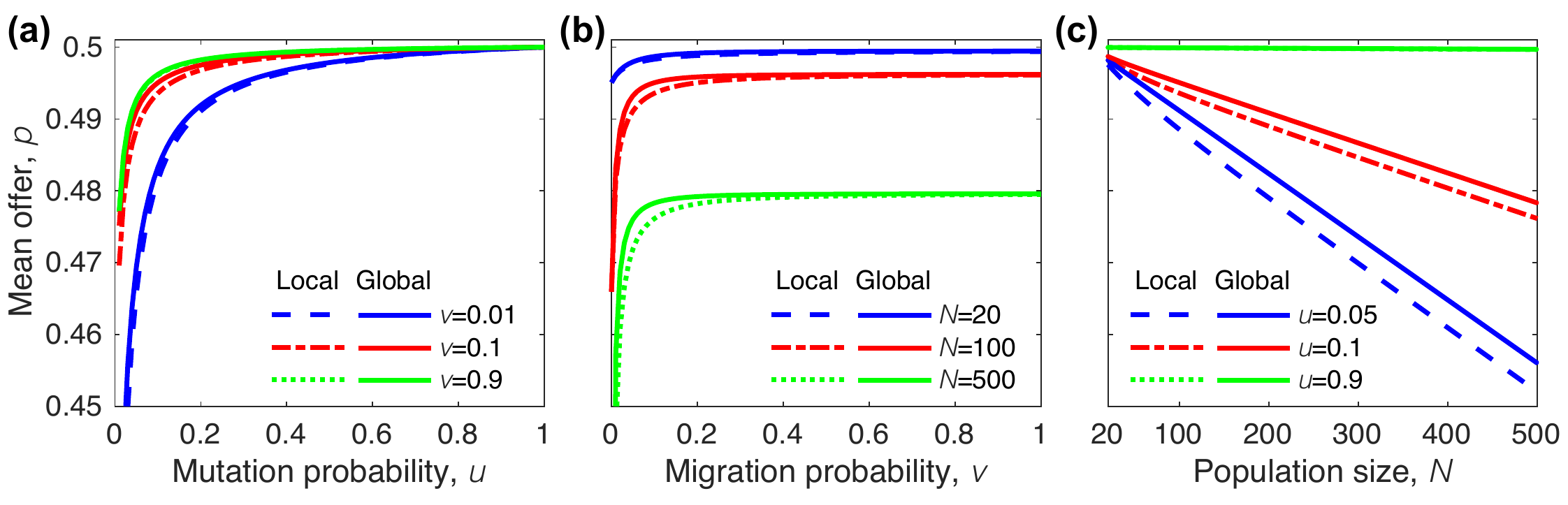}}
\caption{(Color online) The effects of the mutation probability $u$, the migration probability $v$, and the population size $N$ on the mean offer $p$ when all players use empathetic strategies ($\alpha=1$).
$p$ of local migration has a very small difference from $p$ of global migration.
$p$ increases with $u$ in (a), increases with $v$ in (b), but decreases with $N$ in (c).
Parameters: $M = 9$, $\omega=0.001$, (a) $N=100$, (b) $u=0.1$, and (c) $v=0.1$.}
\label{fig5}
\end{figure}
As shown in Fig.~\ref{fig5}, the mean offer of local migration is very close to the one of global migration, implying that the migration pattern has a negligible effect on the evolution of fairness. Therefore, we focus on the mean offer of global migration, which can be obtained analytically. From $f(x)=\frac{1}{M-1}\sum_{j=1}^{M-1}\cos\frac{2\pi j x}{M}$, we have $f(M)=1$ and $f(x)=-\frac{1}{M-1}$ when $x\in\{1,\cdots,M-1\}$. Substituting these values into Eq.~(\ref{empathy}), we have
\begin{eqnarray}
\begin{array}{l}
p=\frac{1}{2} -\omega \frac{(1-u)(N-1)N}{24M}(\frac{1}{1+(N-1)u}+\frac{(M-1)(1-v\frac{M}{M-1})}{(1+(N-1)v\frac{M}{M-1})(1+(N-1)u+(N-1)v\frac{M}{M-1}-(N-1)uv\frac{M}{M-1})}).		\end{array}
\end{eqnarray}
It is easy to verify that the mean offer increases with the mutation probability and the migration probability, but decreases with the population size. The result is also shown in Fig.~\ref{fig5}. Accordingly, mutation and migration promotes the evolution of fairness, but the population size inhibits the evolution of fairness. In order to make sure that almost all players can participate in games, we here assume that the population size is two times more than the group number $M$.

\begin{figure}[t]
\centerline{\includegraphics[width=0.8\textwidth]{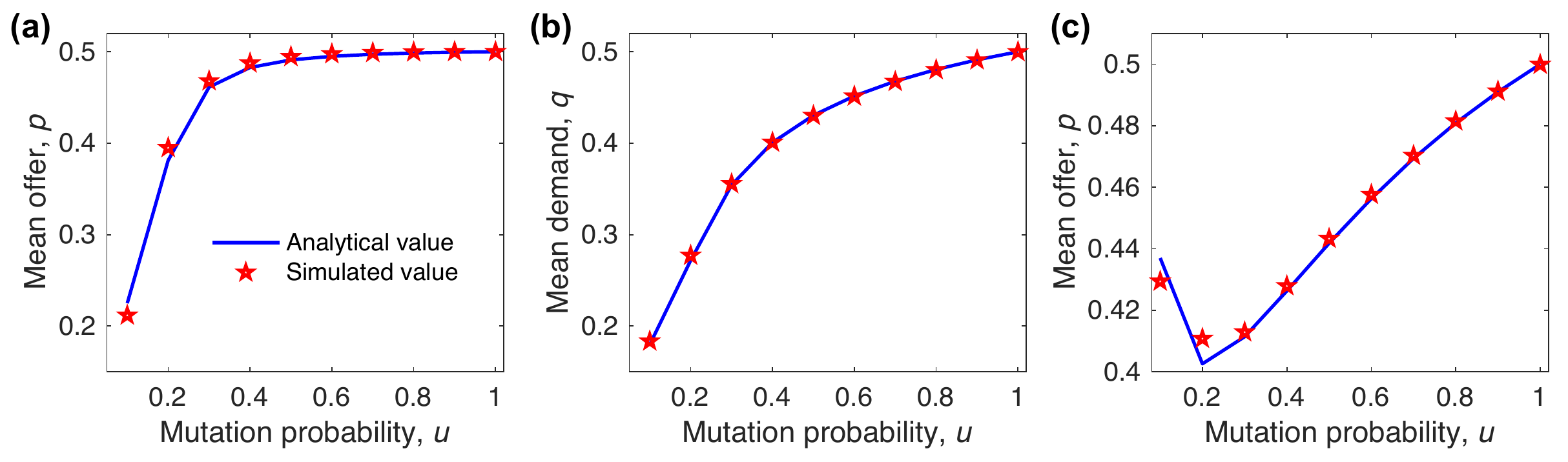}}
\caption{(Color online) Under strong selection ($\omega=1$), comparison of theoretical results and numerical results. ((a), (b)) When all players use independent strategies ($\alpha=0$), our analytical calculations of the mean offer $p$ and demand $q$ ($S=7, 30, 40$) are in good agreement with the simulated values for different mutation probabilities $u$. (c) When all players use empathetic strategies ($\alpha=1$), our analytical calculations of $p$ ($S=100$) are in good agreement with the simulated values for different $u$.
}
\label{fig6}
\end{figure}
It has been shown that under strong selection, the mean offer and demand change significantly with $\alpha$. Moreover, the results for any $\alpha$ are between those under two extreme cases $\alpha=1$ and $\alpha=0$. Therefore, we calculate the approximate values of the mean offer and demand for $\alpha=1$ and $\alpha=0$.
When $\alpha=1$, we use the replicator dynamics of $S$ discrete strategies $A=\{0, \frac{1}{S-1}, \cdots, 1\}$.
By employing $x_i$ to denote the frequency of the strategy $i\in A$, we have
\begin{eqnarray}
\frac{\textit{d} x_i}{\textit{d}t}=\frac{x_i\times\pi_i\times(1-u)}{\phi}+\frac{u}{S}-x_i,
\end{eqnarray}
where $\pi_i$ is the payoff of the strategy $i$, and $\phi=\sum_{i=1}^S x_i \pi_i$ is the average payoff of the population.
When $\alpha=0$, we use the replicator dynamics of $S^2$ discrete strategies.
Similarly, by using $x_{(i,j)}$ to denote the frequency of the strategy $(i,j)\in A\times A$, we have
\begin{eqnarray}
\frac{\textit{d} x_{(i,j)}}{\textit{d}t}  =\frac{x_{(i,j)}\times\pi_{(i,j)}\times(1-u)}{\Phi}+\frac{u}{S^2}-x_{(i,j)},
\end{eqnarray} where $\pi_{(i,j)}$ is the payoff of the strategy $(i,j)$, and $\Phi=\sum_{i=1}^S\sum_{j=1}^S x_{(i,j)} \pi_{(i,j)}$ is the average payoff of the population.
When all strategies have identical initial frequencies, the stationary solutions of the above replicator equations, which no longer change with time, are denoted by $\overline{x}_i$ and $\overline{x}_{(i,j)}$.
When $\alpha=1$, we approximate the mean offer by  $\sum_{i=1}^{S}\frac{i-1}{S-1}\overline{x}_i$ with sufficiently large $S$.
When $\alpha=0$, we approximate the mean offer and the mean demand by $\sum_{i=1}^{S}\frac{i-1}{S-1} \overline{x}_{(i,j)}$ and $\sum_{j=1}^{S}\frac{j-1}{S-1}\overline{x}_{(i,j)}$ with sufficiently large $S^2$, respectively.
As shown in Fig.~\ref{fig6}, our approximations agree well with simulated results for different mutation probabilities. For $\alpha=0$, the mean offer and demand approach to the fair ones with increasing mutation probability, implying that mutation promotes the evolution of fairness.
For $\alpha=1$, an intermediate mutation probability corresponds to the lowest level of fairness (lowest mean offer), implying that mutation may inhibit the evolution of cooperation.
This is different from the counterpart of weak selection, which is that a higher mutation probability leads to a higher level of fairness.

\section{Discussions and Conclusions}
Empathetic players cannot survive in unstructured populations~\cite{page2002empathy}, but can whenever players are restricted to interact with a fraction of the total population~\cite{iranzo2012empathy}.
Moreover, the survival of empathetic players is the precondition of studying the effects of empathy on the evolution of fairness. Accordingly in this paper, we focused on group-structured populations, where interaction occurs among players in the same group.
We introduced empathy by allowing a proportion $\alpha$ of the population to play empathetic strategies.
For high mutation probabilities, the mean offer decreases with $\alpha$ but the mean demand increases with $\alpha$, implying that empathy inhibits the evolution of fairness.
For low mutation probabilities, they both approach to the fair ones with increasing $\alpha$, implying that empathy promotes the evolution of fairness.
The difference of the phenomena above is resulted from the decisive factor of the strategy selection, which is the mutation process for high mutation probabilities and the payoff difference for low mutation probabilities.

We also investigated how the intensity of selection $\omega$ influences the evolution of fairness in finite populations. The mean offer and demand significantly decrease with $\omega$ when it is small, but no longer change with $\omega$ when it is large.
For large $\omega$, the linear dependence of the fitness on the payoff makes $\omega$ omitted from the numerator and the denominator of the probability that a player reproduces an offspring. It implies that sufficiently large $\omega$ no longer enlarges the payoff difference of one strategy over another. Therefore, the mean offer and demand remain around non-zero constants when sufficiently large $\omega$ increases.
The previous study has discussed the problem in unstructured populations~\cite{rand2013evolution}.
It has been shown that sufficiently large $\omega$ drives the mean offer and demand to converge to the rational self-interested strategy $p=q=0$. The difference of this result from our result is caused by the exponential dependence of the fitness on the payoff in the previous model, which implies that the payoff advantage of a lower offer over a higher offer can be increasingly enlarged by raising large $\omega$.

We further analytically calculated the mean offer for an extreme case $\alpha=1$ under weak selection.
Based on this analytical result, we found that mutation and migration can promote the evolution of fairness, while the population size cannot.
The reason why we did not investigate other levels of $\alpha$ is twofold:
a) The mean offer and demand for any $\alpha$ are close to the mean offer for $\alpha=1$; b) The mean offer and demand for $\alpha=0$ have been obtained in the previous literature~\cite{zhang2018effect}, which has only studied the effects of varying the stake on fairness.
Under strong selection, the mean offer and demand for any $\alpha$ are between those for $\alpha=0$ and $\alpha=1$.
Accordingly, we analytically achieved the mean offer and demand for $\alpha=0$ and $\alpha=1$ by using the replicator dynamics, which does not involve population size and migration.
Counterintuitively, we demonstrate that although a higher mutation probability leads
to a higher level of fairness under weak selection for $\alpha=1$, an intermediate mutation probability corresponds to the lowest level of fairness under strong selection.
Our theoretical results above were all verified by numerical simulations.

\section{acknowledgments}
Y.Z. is grateful for the support by the National Natural Science Foundation of China [grant number 61603036, 61520106009, 61533008]. A.L. acknowledges the Human Frontier Science Program Postdoctoral Fellowship [grant number LT000696/2018-C], and the generous support from Foster Lab at Oxford.

\section*{Appendix}
\appendix
When all players use empathetic strategies, the mean offer $p$ is calculated as follows.
We first discretize the continuous strategy $p\in[0,1]$.
Then, we calculate the average frequencies of all discrete strategies according to the results on discrete strategies. When the number of discrete strategies tends to $+\infty$, the weighted average of all discrete strategies whose weights are the frequencies tends to the mean offer.

Assume that all players choose strategies from $S$ discrete empathetic strategies $\{0, \frac{1}{S-1}, \cdots, 1\}$. Now, the original continuous problem is changed into a discrete one, and we focus on the stationary frequency of the above $S$ strategies. According to~\cite{zhang2016impact}, the average frequency of the $k_{th}$ strategy over the stationary distribution under weak selection $(\omega\to0)$, $\langle x_k\rangle_{\omega\to 0}$, is given by
\begin{eqnarray}
\label{xcc1}
\begin{array}{l}
\langle x_{k}\rangle_{\omega\rightarrow 0}
=\frac{1}{S}+\omega\frac{1-u}{Nu}(\Gamma_1(a_{kk}-\overline{a}_{**})+\Gamma_2
(\overline{a}_{k*}-\overline{a}_{*k})+\Gamma_3(\overline{a}_{k*}-\overline{a})),\\
\Gamma_1=(N-1)(N-2)/(3M)\sum_{x=1}^M(-2\Phi_1\Psi_2-\Phi_4\alpha_1+3\Psi_2),\\
\Gamma_2=(N-1)/(3M)\sum_{x=1}^M(3\Psi_1-3\Psi_2 +(N-2)(-2\Phi_1\Psi_2-\Phi_4\alpha_1+\Phi_2\Psi_2+\Phi_3\Psi_1+\Phi_5\alpha_1)),\\
\Gamma_3=(N-1)(N-2)/(3M)\sum_{x=1}^M(3\Psi_1-3\Psi_2+2(2\Phi_1\Psi_2+\Phi_4\alpha_1-\Phi_2\Psi_2-\Phi_3\Psi_1-\Phi_5\alpha_1)),
\end{array}
\label{frequency}
\end{eqnarray}
where $\alpha_1=\frac{1-u}{1+(N-1)u}$, the above $\Phi_i$ and $\Psi_i$ omit $(f(x))$, $\Phi_1(f)=\frac{(1-u)(2-v(1-f))}{2+(N-2)u+\frac{2(N-2)(1-u)v}{3}(1-f)}$, $\Phi_2(f)=\frac{2-u-v(1-f)}{2+\frac{2(N-2)u}{3}+\frac{(N-2)(2-u)v}{3}(1-f)}$,
$\Phi_3(f)=\frac{(1-u)(2-v(1-f))}{2+\frac{2(N-2)u}{3}+\frac{(N-2)(2-u)v}{3}(1-f)}$,
$\Phi_4(f)=\frac{(1-u)(1-v(1-f))}{1+\frac{(N-2)u}{2}+\frac{(N-2)(1-u)v}{3}(1-f)}$,
$\Phi_5(f)=\frac{(2-u)(1-v(1-f)) }{2+\frac{2(N-2)u}{3}+\frac{(N-2)(2-u)v}{3}(1-f)}$,
$\Psi_1(f)=\frac{1-v(1-f)}{1+(N-1)v(1-f)}$, and $\Psi_2(f)=\frac{(1-u)(1-v(1-f))}{1+(N-1)u+(N-1)(1-u)v(1-f)}$.
The omitted $f(x)$ describes the migration pattern.
The expression of $f(x)$ is $f(x)=\frac{1}{M-1}\sum_{j=1}^{M-1}\cos\frac{2\pi j x}{M}$ for global migration and is $f(x)=\cos\frac{2\pi  x}{M}$ for local migration.

The definition of the mean offer $p$ is
$p=\lim_{S\to +\infty}\sum_{k=1}^{S}\frac{k-1}{S-1}\times\langle x_{k}\rangle_{\delta\rightarrow 0}$.
According to Eq.~(\ref{frequency}), the mean offer $\langle p\rangle$ is
\begin{eqnarray}
\begin{array}{l}
p=\lim_{S\to+\infty}\sum_{k=1}^S\frac{k-1}{S-1}\frac{1}{S}+\omega\frac{1-u}{Nu}\lim_{S\to +\infty}\sum_{k=1}^{S}\frac{k-1}{S-1}(\Gamma_1(a_{kk}-\overline{a_{**}}) +\Gamma_2(\overline{a_{k*}}-\overline{a_{*k}})+\Gamma_3(\overline{a_{k*}}-\overline{a})).
\end{array}
\end{eqnarray}
Because
\begin{eqnarray*}
\begin{array}{l}
\lim\limits_{S\to+\infty}\sum_{k=1}^S\frac{k-1}{S-1}\frac{1}{S}=\int_0^1xdx=\frac{1}{2},\\
\lim\limits_{S\to+\infty}\sum_{k=1}^S\frac{k-1}{S-1}\frac{1}{S}a_{kk}=\int_0^1xa(x,x)dx=1/2,\\
\lim\limits_{S\to+\infty}\sum_{k=1}^S\frac{k-1}{S-1}\frac{1}{S}\overline{a_{**}}=\frac{1}{2}\int_0^1a(x,x)dx=1/2,\\
\lim\limits_{S\to+\infty}\sum_{k=1}^S\frac{k-1}{S-1}\frac{1}{S}\overline{a_{*k}}=\int_0^1\int_0^y y^2d{x}d{y}+\int_0^1\int_y^1y(1-x)d{x}d{y}=7/24,\\
\lim\limits_{S\to+\infty}\sum_{k=1}^S\frac{k-1}{S-1}\frac{1}{S}\overline{a_{k*}}=\int_0^1\int_0^yxyd{x}d{y}+\int_0^1\int_y^1x(1-x)d{x}d{y}=5/24,\\
\lim\limits_{S\to+\infty}\sum_{k=1}^S\frac{k-1}{S-1}\frac{1}{S}\overline{a}=\int_0^1\int_0^y y/2d{x}d{y}+\int_0^1\int_y^1(1-x)/2d{x}d{y}=1/4,
\end{array}
\end{eqnarray*}
we have
\begin{eqnarray}
p=\frac{1}{2}-\omega\frac{1-u}{{24Nu}}(2\Gamma_2 + \Gamma_3)=\frac{1}{2}-\omega\frac{(1-u)(N- 1)}{{24Mu}}\sum_{x=1}^M(\Psi_1(f(x)-\Psi_2(f(x)).
\label{empathy}		
\end{eqnarray}


\end{document}